\documentclass[preprint]{aastex}

\usepackage{epsfig}
\usepackage{natbib}
\usepackage{graphicx}    
\usepackage{makeidx}
\makeindex                               
\usepackage{lscape}
\usepackage{verbatim}                                   
\usepackage{amsmath,amsthm,amsfonts,amsopn,amssymb}     
\usepackage{longtable}
\usepackage{pdflscape}
\usepackage{afterpage}
\usepackage{rotating}					                          
\usepackage{ulem}
\usepackage{epstopdf}  
\usepackage{multirow}
\usepackage{color}
\usepackage[left]{lineno}

\usepackage{chngpage}

\received{}
\accepted{}

\slugcomment{Accepted to AJ}
\shorttitle{Planet Hunters 6}
\shortauthors{Schmitt}

\begin{document}
\title{Planet Hunters VI:  An Independent Characterization of KOI-351 and Several Long Period Planet Candidates from the \textit{Kepler} Archival Data\altaffilmark{1}}
\author{
Joseph R. Schmitt\altaffilmark{2},
Ji Wang\altaffilmark{2},
Debra A. Fischer\altaffilmark{2},
Kian J. Jek\altaffilmark{7},
John C. Moriarty\altaffilmark{2},
Tabetha S. Boyajian\altaffilmark{2},
Megan E. Schwamb\altaffilmark{3},
Chris Lintott\altaffilmark{4,5},
Stuart Lynn\altaffilmark{5},
Arfon M. Smith\altaffilmark{5},
Michael Parrish\altaffilmark{5},
Kevin Schawinski\altaffilmark{6},
Robert Simpson\altaffilmark{4},
Daryll LaCourse\altaffilmark{7}, 
Mark R. Omohundro\altaffilmark{7}, 
Troy Winarski\altaffilmark{7}, 
Samuel Jon Goodman\altaffilmark{7}, 
Tony Jebson\altaffilmark{7}, 
Hans Martin Schwengeler\altaffilmark{7}, 
David A. Paterson\altaffilmark{7}, 
Johann Sejpka\altaffilmark{7}, 
Ivan Terentev\altaffilmark{7}, 
Tom Jacobs\altaffilmark{7}, 
Nawar Alsaadi\altaffilmark{7}, 
Robert C. Bailey\altaffilmark{7}, 
Tony Ginman\altaffilmark{7}, 
Pete Granado\altaffilmark{7}, 
Kristoffer Vonstad Guttormsen\altaffilmark{7}, 
Franco Mallia\altaffilmark{7}, 
Alfred L. Papillon\altaffilmark{7}, 
Franco Rossi\altaffilmark{7}, 
and Miguel Socolovsky\altaffilmark{7}} 

\email{joseph.schmitt@yale.edu}

\altaffiltext{1}{This publication has been made possible through the work of more than 280,000 volunteers in the Planet Hunters project, whose contributions are individually acknowledged at \url{http://www.planethunters.org/authors}.  The authors especially thank the Planet Hunters volunteers who participated in identifying and analyzing the candidates presented in this paper. They are individually recognized at \url{http://www.planethunters.org/PH6.}}
\altaffiltext{2}{Department of Astronomy, Yale University, New Haven, CT 06511 USA}
\altaffiltext{3}{Institute of Astronomy and Astrophysics, Academia Sinica, 11F of Astronomy-Mathematics Building, National Taiwan University. No.1, Sec. 4, Roosevelt Rd, Taipei 10617, Taiwan}
\altaffiltext{4}{Oxford Astrophysics, Denys Wilkinson Building, Keble Road, Oxford OX1 3RH}
\altaffiltext{5}{Adler Planetarium, 1300 S. Lake Shore Drive, Chicago, IL 60605, USA}
\altaffiltext{6}{Institute for Astronomy, Department of Physics, ETH Zurich, Wolfgang-Pauli-Strasse 16, CH-8093 Zurich, Switzerland}
\altaffiltext{7}{Planet Hunter}

%

\begin{abstract}

We report the discovery of 14 new transiting planet candidates in the \textit{Kepler} field from the Planet Hunters citizen science program.  None of these candidates overlapped with \textit{Kepler} Objects of Interest (KOIs) at the time of submission.  We report the discovery of one more addition to the six planet candidate system around KOI-351, making it the \textit{only seven planet candidate system} from \textit{Kepler}.  Additionally, KOI-351 bears some resemblance to our own solar system, with the inner five planets ranging from Earth to mini-Neptune radii and the outer planets being gas giants; however, this system is very compact, with all seven planet candidates orbiting $\lesssim 1$~AU from their host star. A Hill stability test and an orbital integration of the system shows that the system is stable.  Furthermore, we significantly add to the population of long period transiting planets; periods range from 124-904 days, eight of them more than one Earth year long.  Seven of these 14 candidates reside in their host star's habitable zone.  

\end{abstract}


\section{Introduction}
\label{intro}

Over the last 20 years, hundreds of exoplanets have been discovered.  One powerful method to discover planet candidates is the photometric transit technique, in which a planet crosses in front of its host star as seen from Earth.  The \textit{Kepler} mission \citep{Borucki2010} has been observing $\sim$160,000 stars nearly continuously for almost four years searching for these transit signals.  In the first 16 quarters, spanning four years, more than 3,800 planet candidates have been discovered (and about 800 more have yet to be dispositioned) via this photometric transit technique\footnote{http://exoplanetarchive.ipac.caltech.edu, last accessed March 11, 2014} \citep{Borucki2011,Batalha2013,Burke2014}.  The \textit{Kepler} team searches for transit signals using a matched filter search algorithm that employs wavelets for numerical efficiency, the transit planet search (TPS) \citep{Jenkins2002,Jenkins2010}, which requires three transits with a significance of $7.1\sigma$ to be placed on the Threshold Crossing Event (TCE) list \citep{Tenenbaum2013,Tenenbaum2014}. TPS's Q1-16 run has discovered more than 16,000 transit signals \citep{Tenenbaum2014}.  Those TCEs which pass additional tests, including a human review stage \citep{Batalha2013}, become \textit{Kepler} Objects of Interest (KOIs).  It is expected that most KOI candidates are true planets \citep{Morton2011}. The false positive rate has been found to depend on the planet radius, with the lowest false positive rate ($6.7-8.8\%$) in the range of $1.25-6.00R_{\oplus}$, where we find a majority of the new Planet Hunters candidates.  Larger planets suffer a false positive rate of $15.9-17.7\%$ \citep{Fressin2013}.  However, the false positive rate for multiple planet candidates is very low, approximately only two out of all \textit{Kepler} targets \citep{Lissauer2012,Lissauer2014}.

The Planet Hunters project\footnote{http://www.planethunters.org/} \citep{Fischer2012} is one of the Zooniverse projects\footnote{https://www.zooniverse.org/} \citep{Lintott2008,Lintott2011,Fortson2012} and is designed to have humans visually check \textit{Kepler} light curves, broken into 30 day increments, to search for undiscovered transit signals.  Since December, 2010, approximately 280,000 public volunteers have searched through more than 21 million \textit{Kepler} light curves hunting for transiting planets, contributing a cumulative total of 200 years of work. 

While Planet Hunters has identified hundreds of transit signals, we only announce candidates that have not been listed as KOI candidates at the time of submission.  Planet Hunters has discovered more than 40 new planet candidates \citep{Fischer2012,Lintott2013,Wang2013a,Schwamb2013}, including two confirmed planets. The first confirmed planet from the Planet Hunters project is PH1~b (Kepler-64b), a circumbinary planet in a $\sim$137 day orbit around an eclipsing binary, and is the first known planet in a quadruple star system \citep{Schwamb2013}.  The second confirmed planet from the Planet Hunters project is PH2~b (Kepler-86b), a gas giant planet residing in its host star's habitable zone \citep{Wang2013a}. Statistical completeness analysis within the Planet Hunters project is performed by injecting fake transit events into real \textit{Kepler} light curves \citep{Schwamb2012}. This analysis shows that Planet Hunters are effective at detecting transits of Neptune-sized planets or larger ($\gtrsim85\%$ completeness for short periods, $P<15$ days), although smaller planets can still be recovered.

In this paper, we present a total of 14 new candidates from Planet Hunters project, all with period greater than 124 days, eight of them more than 1 Earth year long.  Additionally, seven of the new candidates lie within the most recent HZ estimates \citep{Kopparapu2013}. Another candidate discovered around KIC 6436029 makes this a multiple planet candidate system. The new planet candidate we detect in the known KOI six candidate system, KOI-351, makes it the only \textit{Kepler} star with a seven planet candidate system.  While this paper was in the review process, \citet{Cabrera2014} and \citet{Lissauer2014} independently discovered and characterized the seventh candidate in KOI-351.  Also while in the review process, six of our new candidates were classified as candidates (KIC 2437209) or ``NOT DISPOSITIONED'' (KIC 5094412, 6372194, 6436029, 6805414, 11152511) KOIs on the Exoplanet Archive (accessed March 11, 2014), all of which were detected as TCEs in the latest TPS Q1-16 search\citep{Tenenbaum2014}.  Of the new candidates with 3+ transits in Q1-16, all except the seventh candidate in KOI-351 have now been detected by TPS.  Section~\ref{discovery} explains how these new planet candidates were discovered. Section ~\ref{transitchar} explains our method to calculate the transit parameters, stellar parameters, whether the planet resides in the HZ, and a discussion of the false positive tests that we have carried out for our new planet candidates.  Section~\ref{newcand} discusses characteristics of notable new candidates.  We conclude in Section~\ref{conc}.

\section{Planet Hunters Candidate Discoveries}
\label{discovery}

Candidates are identified through one of two ways.  The classic method is through Planet Hunters interface \citep{Fischer2012,Schwamb2012}, in which users are shown a 30-day light curve and asked to identify transit-like features. The Planet Hunters interface has shown quarters 1, 2, part of 3, 4, 5, 7, 9, 14, and part of 16.  However, once a planet candidate is identified, volunteers typically search through all available quarters for additional transits of the same planet candidate or for transits of new candidates within the same system.  We have implemented a weighting scheme \citep{Schwamb2012} in order to rank the quality of user transit classifications. In brief, all users start out with an equal weight, and synthetic light curves are used to seed the user weighting. Users who properly identify synthetic transits are given higher weights. The user weightings continue to evolve depending on whether or not individual rankings agree with the majority rankings. Transits above a threshold score are then sent to the science team to be analyzed. 

The other way candidates are identified is via the Planet Hunters Talk page\footnote{http://talk.planethunters.org} \citep{Lintott2013,Wang2014}.  This discussion tool allows users to publicly post and discuss interesting light curves with others. It is through this interface that users are easily able to download all data and collectively scrutinize potentially interesting light curves.  The interface provides quick links to the MAST\footnote{http://archive.stsci.edu/kepler/} and UKIRT databases for each object.  Publicly available web-based tools hosted at the NASA Exoplanet Archive (NEA) and SkyView\footnote{http://skyview.gsfc.nasa.gov/} are used frequently to calculate periodograms, normalize, and phase-fold the data, as well as performing data validation to rule out false positives. Planet Hunters volunteers are \textit{instrumental} in the success of the Planet Hunters project. 

The Planet Hunter volunteers organize different types of light curves in collections on the Planet Hunters Talk page. Once a collection has been established, users can compare the light curves they are examining with the light curves in the collections. If the user decides to discuss the light curve on the Talk page, they can suggest that the light curve be added to an existing collection.  The Planet Hunters science team searches the collections for interesting candidates. Alternatively, some of the active users compile lists of prospective candidates and pass the spreadsheets to the science team for further vetting. The science team then carries out data validations tests described in Section ~\ref{FP}.  This paper reflects discoveries made through the Planet Hunters Talk interface.

\section{Transit Characterization}
\label{transitchar}

\subsection{Data Validation} 
\label{FP}

Once a system is identified as having a possible transit signal, we perform a full analysis of the system using the \textit{Kepler} light curve (data validation; \citealt{Batalha2010}) and any other publicly available archival data.  This includes using the \text{PyKE} package \citep{Still2012} and screening the surrounding field for background eclipsing binaries (BGEB) by ensuring that the flux-weighted centroid remains stationary during an object's transit \citep{Bryson2013}.  By measuring the flux centroid shift between in- and out-of-transit, a limit can be put on the angular separation of a possible contamination  source. This method has limitations when the target star is brighter than 11th magnitude or if it is in a crowded field \citep{Wang2014}. The flux-weighted centroid method reported large ($>3\sigma$) flux centroid offsets for the aforementioned cases even for confirmed planet candidates. However, all but one objects in this paper are fainter than 13th magnitude. Our own flux-weighted centroid analysis package is described in \citet{Wang2013a} and \citet{Wang2014}.

\begin{figure*}[tp]
	\centering
		\centerline{\includegraphics[width=1.00\textwidth]{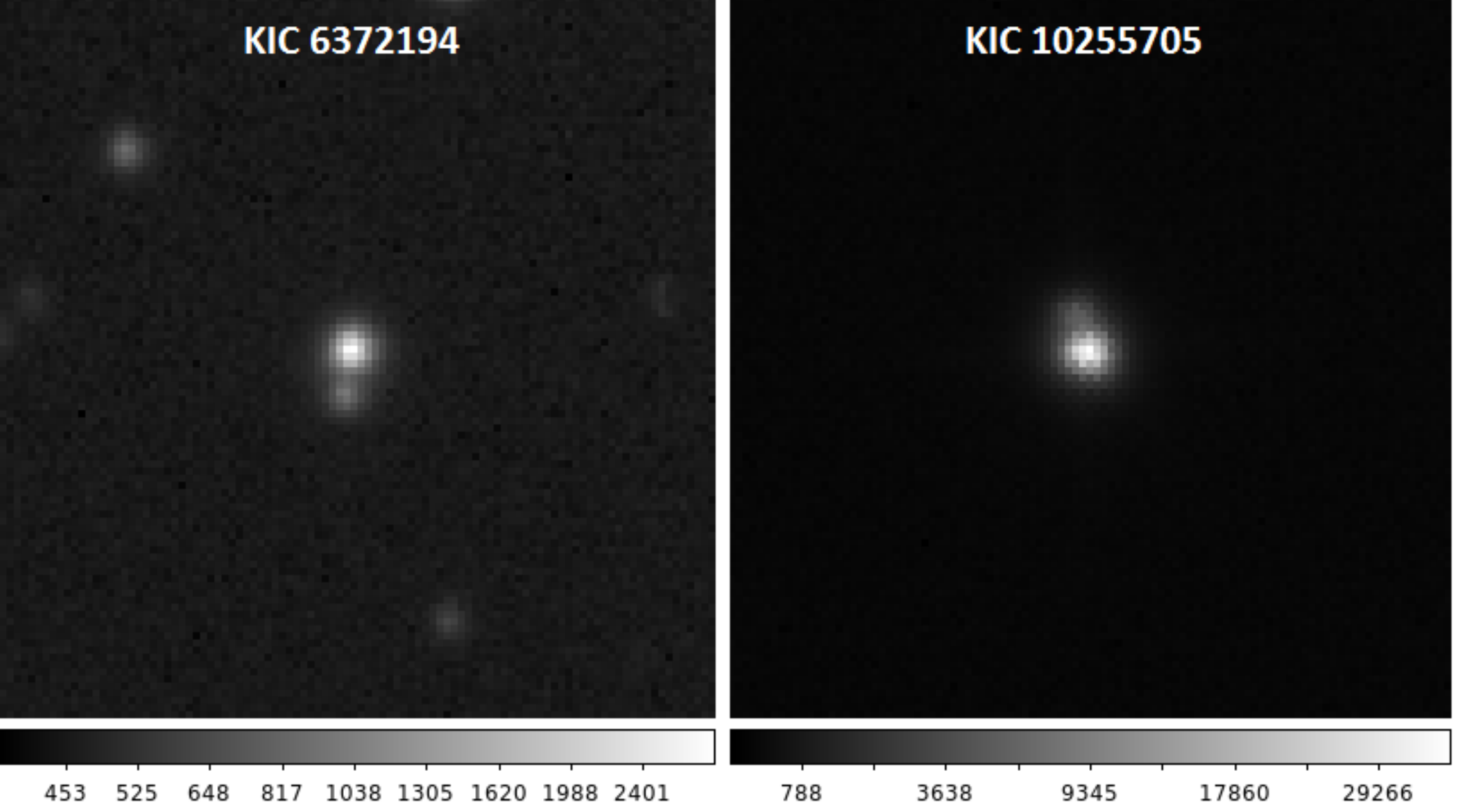}}
	\caption{UKIRT images of the stars with companions within 4\arcsec.  The images are 20\arcsec on a side  Only KIC 10255705 has a companion within its $3\sigma$ confusion radius.}
	\label{fig:AO}
\end{figure*}

We use UKIRT and 2MASS images to search for nearby contaminating sources and asymmetric point spread functions, necessary for probing nearby unresolved sources due to the low-resolution of the \textit{Kepler} CCD detector. Two of the stars have a visual companion within 4\arcsec:  KIC 6372194, and 10255705 (see Figure~\ref{fig:AO} for the UKIRT images and Table~\ref{tab:AO}).  If these companions are outside the computed confusion radius, the candidates are unlikely to be orbiting the contaminating sources. If they are orbiting the contaminating sources and outside the confusion radius, an apparent pixel centroid offset should have been detected. However, KIC 10255705 has a $3\sigma$ confusion radius of $3.2\arcsec$ and a companion at $1.4\arcsec$, meaning that no apparent pixel centroid offset would be seen, so we cannot be sure the planet is truly orbiting the brighter star.  The neighboring stars for both KIC 6372194 and 10255705 are at least two magnitudes fainter than the central star.  The uncertainty caused by the flux contamination is smaller than the one from the stellar radius estimation, so we did not include the dilution factor in our parameter estimates.  Afterwards, the light curves are then fully modeled and inspected for variations in even-odd transit depth, the presence of secondary eclipses, and misshaped transit profiles, any of which can indicate a falsely identified planetary candidate with tools from \citet{Still2012}.

\subsection{Stellar Properties}
\label{sec:stellarchar}

The revised set of \textit{Kepler} stellar parameters provides estimates of $T_{\rm eff}$, $\log{g}$, mass, radius, and [Fe/H] of all stars in the \textit{Kepler} field \citep{Huber2014}. We adopt these as stellar inputs for the \citet{Wang2013a} transit fitting routine from the \textit{Kepler} stellar data hosted by the NEA Exoplanet Archive\footnote{http://exoplanetarchive.ipac.caltech.edu/}.  Table~\ref{tab:st} lists the primary stellar parameters.  We use the values in Table~\ref{tab:st} as inputs for our own transit fitting routine \citep{Wang2013a,Wang2014} developed to iterate between the light curve solution and the Yonsei-Yale (Y$^2$) isochrones \citep{Demarque2004} for a range of ages spanning from $0.08-15$~Gyr, with [$\alpha$/Fe]~$=0$. The high $T_{\rm{eff}}$ of KIC 5522786, $T=8941$ K, fell on the upper edge of the temperature range for our stellar models, which caused error bars to be clipped.  As such, we retain the \citet{Huber2014} stellar parameters for KIC 5522786.  We ran 1000 trials of Monte Carlo simulations using the Y$^2$ Isochrones to obtain stellar properties such as effective temperature, mass, radius, metallicity, and luminosity. The distributions of the inputs are assumed to follow a Gaussian function. The mean and standard deviation are the reported value and error bar in the \textit{Kepler} stellar parameter table. The distributions of the outputs are used to constrain the transit light curve fitting, excluding mathematically acceptable fits that are physically unlikely.  The formal error bars sometimes result in unrealistically low error bars for $\log{g}$, radius, and mass, so we adopt a floor on the $\log{g}$ error bar of $\pm0.10$, 20\% for the radius error, and 10\% for the mass error. The outputs of the transiting light curve fitting (e.g., stellar density) can also be used to constrain the Y$^2$ isochrone. Thus, the iterative transit fitting routine is able to provide a set of self-consistent stellar and orbital solutions. 

\subsection{Transit Fitting}
\label{sec:transitfit}

All available long-cadence light curves from \textit{Kepler} quarters 1-16 were flattened, normalized, and phase-folded using the PyKE package \citep{Still2012}.   We used a custom-made package \citep{Wang2013a} to find the best fit values for the orbital period ($P$), the ratio of the planet radius to the stellar radius ($R_{\rm{PL}}/R_{*}$), the ratio of the semi-major axis to the stellar radius ($a/R_*$), inclination ($i$), eccentricity ($e$), longitude of periastron ($\omega$), and midtransit times. Quadratic limb darkening parameters are determined by interpolating a table provided by~\citet{Claret2011}. The best fit parameters were determined through a Levenberg-Marquardt least square algorithm, while the error bars are estimated with a bootstrapping method in the following way. We repeatedly fit the simulated transiting light curves, which were generated from the observations but perturbed by photon noise. In order to reduce the dependence of the initial guess of orbital parameters, we perturbed the initial guess based on the standard deviation of previous runs. We used a range of five times of standard deviation to explore a large phase space. The reported orbital solutions in Table~\ref{tab:pl} are the weighted averages based on the goodness of fit.

\begin{figure*}[tp]
	\centering
		\centerline{\includegraphics{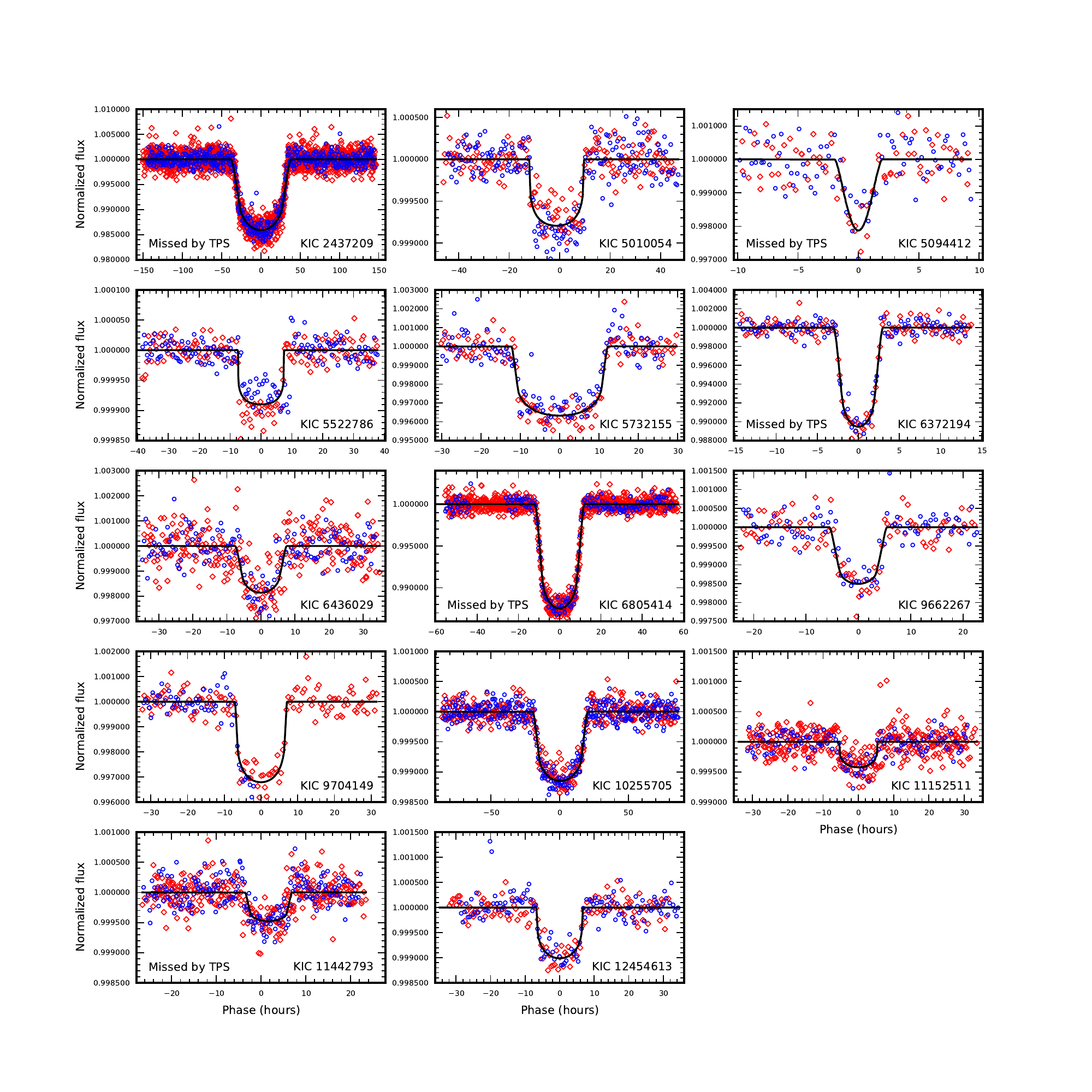}}
	\caption{Phase-folded transit models for new Planet Hunters candidates.  Blue circles represent odd transits and are overplotted onto the red diamonds, which represent even transits.  There are no significant differences between the depths of the even and odd transits for any candidate. See Section~\ref{sec:transitfit} for details.}
	\label{tile}
\end{figure*}

The phase-folded solutions for each star are shown in Figure~\ref{tile}.  Odd and even transits are colored blue and red, respectively, and there is no significant odd-even depth variation, which would be indicative of an eclipsing binary star system.  The strongest odd-even depth variation is $1.49\sigma$ for KIC 5522786, a two transit candidate.  Table ~\ref{tab:pl} contains best fit parameters for the new planet candidates.

\subsection{Habitable Zone}

\begin{figure*}[tbp]
	\centering
		\includegraphics[width=1.00\textwidth]{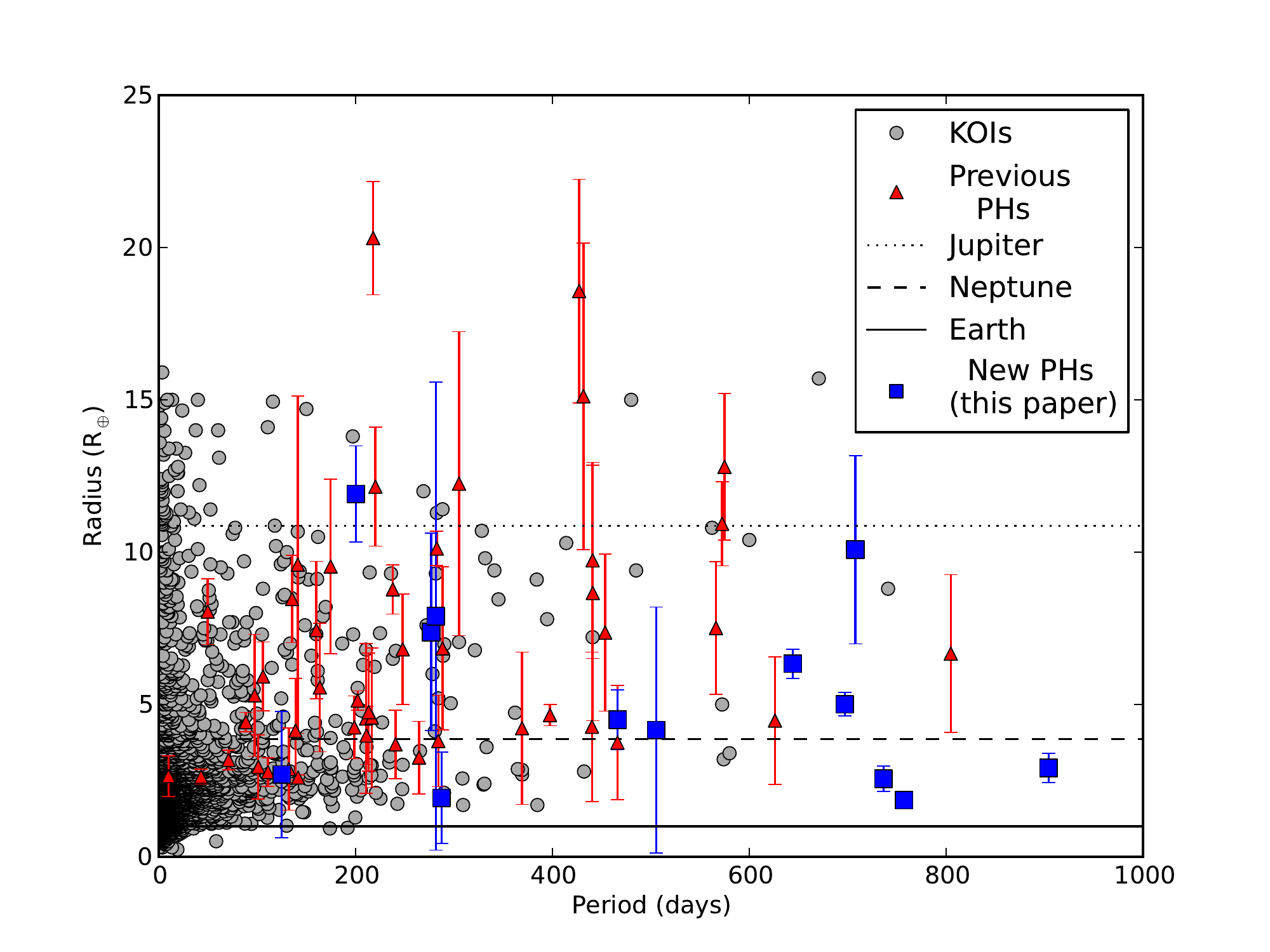}
	\caption{Radius vs. period for the new Planet Hunters candidates (blue squares), previous Planet Hunters candidates or discoveries (red triangles) \citep{Fischer2012,Lintott2013,Wang2014}, and the KOI candidates as of February 6, 2014 (grey circles).  	Planet Hunters provide a large number of the known, transiting long period candidates.  The candidate around KIC 2437209 ($P=281.329$ days, $R=62.64\pm19.61R_{\oplus}$) is excluded for convenience.}
	\label{pvr}
\end{figure*}

Planet Hunters has produced a relatively large proportion of the known long period candidates from the \textit{Kepler} data (see Figure~\ref{pvr}).  Importantly, it is these longer periods that probe the HZs of G- and K-type stars.  The inner edge of the HZ is typically defined as the orbital radius of the water-loss limit, the point at which a terrestrial planet will quickly lose its water, and the outer edge is defined by a maximum greenhouse effect from CO$_2$ \citep{Kasting1993}.  These two boundaries define the ``conservative'' HZ.   ``Empirical'' HZ estimates expand the HZ by assuming Venus had water for much of its history (``Recent Venus'') and that Mars had water early in its history (``Early Mars'').  A revised estimate of the HZ by \citet{Kopparapu2013} proposes to define the HZ by its level of incident flux, $S$, rather than the equilibrium temperature of the planet, removing the dependency on Bond albedo.  The HZ also varies with spectral type as the peak wavelength of the stellar emission changes.  As such, cooler stars have slightly more distant HZs relative to the incident flux on the planet.  

\begin{figure*}[tbp]
	\centering
		\includegraphics[width=1.00\textwidth]{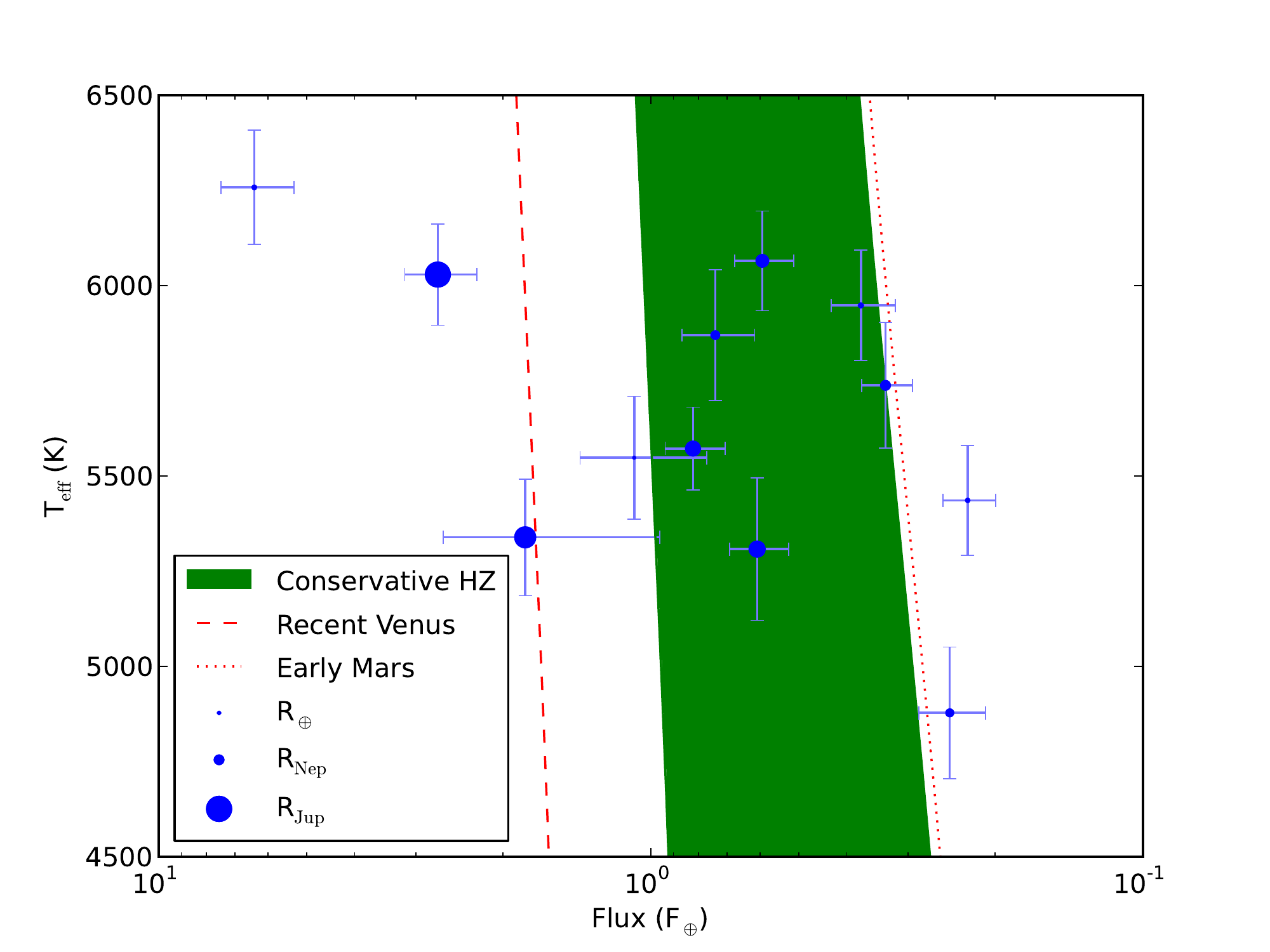}
	\caption{Locations of the new Planet Hunters candidates relative to their host star's HZ.  The green region is the conservative HZ, and the red dashed and dotted lines are the Recent Venus and Early Mars HZ edges, respectively.  The size of the symbol is directly proportional to the physical size of the planet. Seven planet candidates lie directly in the conservative HZ: KIC 2437209, 5010054, 5094412, 5732155, 6372194, 9662267, and 9704149.  KIC 11152511 sits in the Recent Venus HZ with error bars into the conservative HZ. KIC 6436029 and KIC 10255705 are outside of the HZ, but are within 1σ of the Early Mars and Recent Venus HZs, respectively.  KIC 5522786 is too hot for the calibration in \citet{Kopparapu2013}, but at $5S_0$, it is certainly too hot to be habitable.  The candidate around KIC 2437209 ($T_{\rm{eff}}=4842$ K, $S=14.39\pm11.87S_0$) is excluded for convenience.}
	\label{hz}
\end{figure*}

In this revised HZ estimate, only the $a/R_{*}$ and $T_{\rm{eff}}$ are required to determine whether the planet resides in its host's HZ.  Seven planet candidates lie directly in the conservative HZ: KIC 2437209, 5010054, 5094412, 5732155, 6372194, 9662267, and 9704149.  KIC 11152511 sits in the Recent Venus HZ with error bars into the conservative HZ. KIC 6436029 and KIC 10255705 are outside of the HZ, but are within 1σ of the Early Mars and Recent Venus HZs, respectively.  See Figure~\ref{hz} for the location of each planet candidate relative to its host star's HZ.  

Of these, KIC 11152511 has the lowest radius, $R=1.93\pm1.50R_{\oplus}$.  The best-fit value falls very near or on the transition zone between high density super Earths with thin atmospheres and low density mini-Neptunes with thick atmospheres, which recent studies show is likely between 1.5 and 2.0~$R_{\oplus}$ \citep{Lopez2013,Weiss2014,Marcy2014}.  However, this candidate is likely shrunk due to neglected light dilution from its neighboring star.  KIC 5522786's planet candidate also lies in this transition zone ($R=1.86\pm0.25R_{\oplus}$), leaving uncertainty as to whether this candidate would be a super Earth or mini-Neptune.  However, although it lies outside the $T_{\rm{eff}}$ range studied by \citet{Kopparapu2013}, it certainly orbits too close to the star, with $S\approx5S_0$ (five times the solar incident flux).

\section{New Planet Candidates}
\label{newcand}

The best-fit parameters for the 14 new planet candidates in this paper are listed in Table ~\ref{tab:pl}. Their periods and radii are plotted in Figure ~\ref{pvr}.  Comments on individual candidates are given below.

\subsection{KIC 2437209}
\label{2437209}

In quarters $1-16$, the light curve for KIC~2437209 has four large ($\sim 14000$ ppm) and long transits (duration of 71 hours; see Figure~\ref{tile}, Table~\ref{tab:pl}).  Although there were three transits (and a fourth in a data gap), one transit was de-emphasized due to a sudden pixel sensitivity dropout detector in TPS, leaving it without the requisite three detected transits to be labeled a TCE (J. M. Jenkins 2014, private communication). With a fourth transit in Q14, the Q1-16 TPS search \citep{Tenenbaum2014} has detected this signal, and it is now a KOI candidate.  

KIC~2437209's transit duration, approximately equivalent to the duration of Neptune transiting the Sun, is extremely long for a 281 day orbit.  Our first attempts at modeling this system resulted in a degeneracy in the stellar and planetary radius.  This system was consistent with two scenarios: an evolved star on the giant branch with a stellar companion in a circular orbit or with a main sequence dwarf with a planetary companion in a highly eccentric $e=0.974$ orbit viewed at apastron, an unfavorable probability of occurrence.  In the evolved star interpretation, the stellar parameters were $\log{g}=3.07^{+0.60}_{-0.10}$, $R_{*}=4.97^{+2.09}_{-1.14}R_{\odot}$, $R_{\rm{PL}}=62.64\pm19.61R_{\oplus}$, $e=0.43^{+0.23}_{-0.43}$, while the dwarf interpretation had best-fit values of $\rm{log(g)}=4.56\pm0.10$, $R^{*}=0.80\pm0.10R_{\odot}$, $R_{PL}=9.79\pm0.62R_{\oplus}$, $e=0.974^{+0.013}_{-0.029}$,  similar to the \citet{Huber2014} stellar parameters.  

This degeneracy led us to obtain one spectrum of the star with Keck/HIRES \citep{Vogt1994} in an attempt to better determine the stellar parameters.  We used Spectroscopy Made Easy (SME) \citep{Valenti1996,Valenti2005} to model the spectrum; however, due to the faintness of the star ($K_p = 16.353$), the resulting signal to noise ratio was $\sim15$, well below the preferred regime for reliable SME analysis.  Nevertheless, our results fitting the spectrum with SME gave $T_{\rm{eff}}=4965\pm 100$ K, $\log{g}=3.75\pm 0.2$, and $\rm{[Fe/H]}=0.564\pm 0.10$.  This has overlapping error bars with our evolved star interpretation and is inconsistent with input stellar parameters from \citet{Huber2014}.  

KIC~2437209 has six quarters of short-cadence data and likely belongs to the open cluster NGC~6791.  With $g'=17.239$ and $(g'-r')=1.001$ given by the \textit{Kepler Input Catalog}, KIC~2437209 is placed on the giant branch of NGC~6791 on the $g'$ vs. $(g'-r')$ color-magnitude diagram \citep{Platais2011}.  Evolved cluster stars are usually targets for short-cadence \textit{Kepler} observations to study pulsations.  Assuming that the star is a member of NGC~6791, the stellar parameters become $T_{\rm{eff}}=4664\pm100$ K, $R=3.89\pm0.22R_{\oplus}$, and $\log{g}=3.34$ (D. Huber 2014, private communication), which is consistent with our evolved star fit.  Therefore, KIC~2437209 is very likely to be a giant star with a smaller stellar companion.  However, should this planet indeed have $e=0.974$, then this planet may be one of the super-eccentric planets predicted in \citet{Socrates2012} caught in the act of high eccentricity migration.

\subsection{KIC 5010054}

Planet Hunters volunteers discovered an additional transit signal in Q16 at 2496333.91 JD, a signal consistent with a second planet in the system.  No accompanying transit is seen.  	A \texttt{TAP} analysis \citep{Carter2009,Gazak2012,Eastman2013} suggests a period of $P=504^{+170}_{-160}$ days, meaning an earlier transit may have been located in a data gap.  The \texttt{TAP} fit with the modeled stellar parameters gives this signal a radius of $R=2.9\pm0.48R_{\oplus}$.  

\subsection{KIC 5094412}
\label{kic5094412}

KIC~5094412 has the shortest transit duration of candidates in this paper (3.77 hours), leading to few in-transit data points.  We note that two of the four transits appear V-shaped, a shape that can be produced by a grazing eclipsing binary system, but one is an even numbered transit and the other is an odd numbered transit.  The transit morphology of this system is unclear due to the under-sampling induced by the short duration transits. This is the only candidate in our sample that has such a potential V-shaped transit profile.  With three transits in Q1-12 (and one in the data gap between Q6 and Q7), one transit was de-emphasized for a reaction wheel zero crossing due to possible false alarms caused by such events.  It has since been shown that a reaction wheel zero crossing does not cause false alarms, so they are no longer de-emphasized (J. M. Jenkins 2014, private communication).  This change and an extra transit in Q14 led to a detection by TPS in the latest run.  As such, it is now an undispositioned KOI.  

\subsection{KIC 5522786}

KIC 5522786 has only two transits and both exhibit a very sharp ingress and egress. This is the second smallest candidate in this paper, a planet on the super-Earth/mini-Neptune transition zone at $R=1.86 \pm 0.25R_{\oplus}$.  Despite a depth of only $90 \pm 13$ ppm, the transits are visually apparent due to the host star's brightness, a Kepler-magnitude $= 9.350$ star.  A third transit of this candidate would be extremely valuable and is expected to take place at about 2457387.7614 JD (December 31, 2015).  We also note that the star has the highest effective temperature in our sample with $T_{\rm{eff}} = 8941^{+258}_{-396}$ K.  

\subsection{KIC 6372194} 

KIC 6372194 was undetected in the Q1-12 TPS search due to a bug in the Robust Statistic code (J. M. Jenkins 2014, private communication).  The star was not observed until Q4.  However, the bug resulted in the code checking the expected transit in Q3, for which KIC 6372194 was not observed.  As such, it was rejected.  This bug has since been corrected, and the signal is now detected and is designated an undispositioned KOI. 

\subsection{KIC 6436029}

KIC 6436029 already has one KOI candidate (KOI~2828.01; $P=59.5 d$, $R=4.1\pm2.0R_{\oplus}$).  We have detected three transits of another planet around this star with $P=505.45$ days.  This new planet candidate possibly lies in the outer HZ; its upper error bar overlaps the Early Mars zone of the optimistic HZ.  It has a radius of $R=4.16\pm4.04R_{\oplus}$.

\subsection{KIC 6805414}

There are four transits in Q1-12 (a fifth was in the data gap between Q9 and Q10).  Two of these were de-emphasized due to the TPS sudden pixel sensitivity dropout detector (Jon M. Jenkins 2014, private communication).  It was subsequently detected in the Q1-16 search and is now an undispositioned KOI. This candidate was also independently discovered in \citet{Huang2013}.  

\subsection{KIC 11442793 (Kepler-90)}

KIC 11442793 has six listed KOI candidates (KOI-351) at periods of 7.01, 8.72, 59.74, 91.94, 210.60, and 331.64 days (see Table ~\ref{tab:351} for transit parameters).  Planet Hunters has detected an additional 124.92 day signal with five full transits and two partial transits.  However, after the first transit (Q2), the next transit fell within a data gap.  The following transit contained only the egress due to a data gap within Q5.  In the next transit at the end of Q6, only the ingress is observed.  The next transit fell into a data gap.  Two other transits were de-emphasized due to the aforementioned reaction wheel zero crossings.  Furthermore, the large number of candidates in the system caused the data validation to time out.  All these conspired together and led to its non-detection in the Q1-12 TPS search (J. M. Jenkins 2014, private communication).  While this paper was being reviewed, \citet{Cabrera2014} and \citet{Lissauer2014} independently discovered the 124.92 day signal and characterized all seven candidates claiming confirmation, with \citet{Lissauer2014} statistically confirming all seven with high confidence.  For the 124.92 day signal described in this paper, both of the previous works agree well with each other and with our analysis.   

To test whether the seventh signal is merely an alias of one or more of the previous six candidates, we assumed a linear ephemeris for all planets and calculated each planet's midtransit points (see Table ~\ref{tab:ephem}).  None of the planets matched the seventh candidate.  In fact, somewhat surprisingly, \textit{only one} midtransit of the previously known six planet candidates comes within 24 hours of a midtransit of the seventh candidate, but unsurprisingly, it's the candidate with the shortest period (7.01 days).  However, we recognize the fact that we cannot assume a perfectly linear ephemeris on account of transit timing variations, or TTVs \citep{Miralda2002,Agol2005,Holman2005,Holman2010}.  

\begin{figure*}[tp]
	\centering
		\centerline{\includegraphics{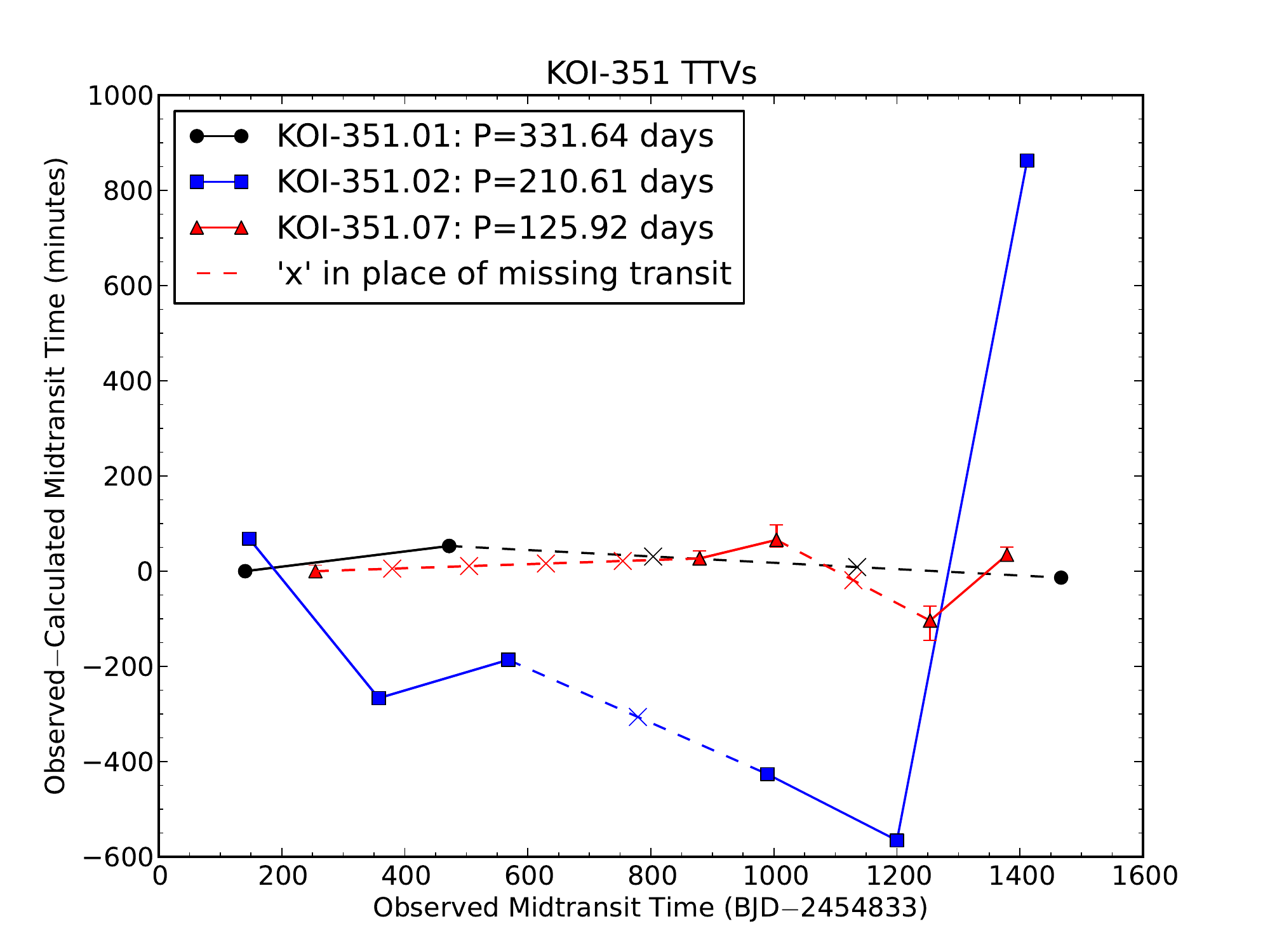}}
	\caption{Observed$-$calculated ($O-C$) midtransit times for the outer three planet candidates of KOI-351: KOI-351.01 (black circles), KOI-351.02 (blue squares), and KOI-351.07 (red triangles).  An `x' corresponds to the approximate time a transit was expected to occur, but was missed due to a data gap. Solid lines are used to connect consecutive observed transits, while dashed lines are used to connect non-consecutive observed transits.  Error bars are plotted and are smaller than the markers where not seen.  The $O-C$ of the last midtransit of KOI-351.02 is $\sim24$ hours larger than the $O-C$ of the second-to-last midtransit of KOI-351.02, making KOI-351.02's TTVs one of the largest known \citep{Mazeh2013}.  The presence of TTVs increases the likelihood that the planets are real and within the same system.}
	\label{koi351ttv}
\end{figure*}

We therefore analyzed the system using the IDL program \texttt{TAP} to determine transit midpoints and compared the observed midtransits to the midtransits expected from the KOI epochs and periods.  A plot of the observed midtransit times minus calculated midtransit times ($O-C$) shows deviations from a perfectly linear ephemeris due to TTVs for the outer three planets, indicating significant gravitational interactions between planets.  Figure~\ref{koi351ttv} shows the $O-C$ plot for the outer three planet candidates of KOI-351. The last transit of KOI-351.02 has $O-C=+14.4$ hours, while the second-to-last transit has $O-C=-9.4$ hours, a change of 24 hours over the course of one orbit.  KOI-351.02 has one of the largest TTVs known, excluding circumbinary planets \citep{Mazeh2013}.  The presence of TTVs significantly increases the likelihood that our planet candidates are real, as they show mutual gravitational interactions with their neighbors.  A more in-depth analysis can likely be used to determine masses and confirm several planets in this system, but this is left for future studies.

The TTVs show that we cannot rely solely on comparisons of linear ephemerides.  Therefore, we also compared the transit depths and transit durations of all seven candidates.  The transit depth for KOI-351.07, $480\pm87$ ppm, is only consistent with two of the KOI candidates, KOI-351.03 and KOI-351.04, but the duration of the seventh candidate is hours longer than either.  These two candidates also have \textit{significantly} different midtransit times and thus cannot be the same object as KOI-351.07.  The depths and durations are compared in Table ~\ref{tab:ephem}, which also shows the phase information of each KOI-351.07 transit relative to the other six candidates.  If KOI-351.07 were in fact a secondary eclipse from one of the other six objects, we would see a constant phase offset of KOI-351's transit relative to the primary eclipses. No such phase offset is observed.  A combination of looking at the differences in duration, depth, midtransit points, and phases of KOI-351.07 relative to the other six candidates clearly demonstrates that KOI-351.07 is a distinct object.  Simply put, the seventh candidate is not an alias of any of the other six candidates.

We also performed two stability tests to assess the feasibility of this system.  Hill stability is a simple stability diagnostic that can be used to determine whether this somewhat compact system is at least feasible.  To test the stability of the system, the masses of the planets must be assumed.  For the inner five planets, which all have $R<4.0R_{\oplus}$, we used the recent \citet{Weiss2014} mass-radius relationship shown below based on the masses and radii of confirmed planets, which was valid for $1.5R_{\oplus} \leq R \leq 4.0R_{\oplus}$.  The middle three planets fall into this range, while the inner two are within $1\sigma$ of this range.  
We also performed two stability tests to assess the feasibility of this system.  Hill stability is a simple stability diagnostic that can be used to determine whether this somewhat compact system is at least feasible.  To test the stability of the system, the masses of the planets must be assumed.  For the inner five planets, which all have $R<4.0R_{\oplus}$, we used the recent \citet{Weiss2014} mass-radius relationship shown below based on the masses and radii of confirmed planets, which was valid for $1.5R_{\oplus} \leq R \leq 4.0R_{\oplus}$.  The middle three planets fall into this range, while the inner two are within $1\sigma$ of this range.  
We also performed two stability tests to assess the feasibility of this system.  Hill stability is a simple stability diagnostic that can be used to determine whether this somewhat compact system is at least feasible.  To test the stability of the system, the masses of the planets must be assumed.  For the inner five planets, which all have $R<4.0R_{\oplus}$, we used the recent \citet{Weiss2014} mass-radius relationship shown below based on the masses and radii of confirmed planets, which was valid for $1.5R_{\oplus} \leq R \leq 4.0R_{\oplus}$.  The middle three planets fall into this range, while the inner two are within $1\sigma$ of this range.  
We also performed two stability tests to assess the feasibility of this system.  Hill stability is a simple stability diagnostic that can be used to determine whether this somewhat compact system is at least feasible.  To test the stability of the system, the masses of the planets must be assumed.  For the inner five planets, which all have $R<4.0R_{\oplus}$, we used the recent \citet{Weiss2014} mass-radius relationship shown below based on the masses and radii of confirmed planets, which was valid for $1.5R_{\oplus} \leq R \leq 4.0R_{\oplus}$.  The middle three planets fall into this range, while the inner two are within $1\sigma$ of this range.  

\begin{equation}
\frac{M_P}{M_{\oplus}}= 2.69 \left(\frac{R_P}{R_{\oplus}}\right)^{0.93}, R<4R_{\oplus}
\end{equation} 

The masses of the upper two planets were calculated with the same equation used by \citet{Lissauer2011b}, shown below.  This equation was determined by fitting for the solar system planets Earth, Uranus, Neptune, and Saturn.  However, the upper error bar on the outermost planet is consistent with a Jupiter radius where the mass-radius relationship breaks down, meaning the mass could be significantly larger.

\begin{equation}
\frac{M_P}{M_{\oplus}}= \left(\frac{R_P}{R_{\oplus}}\right)^{2.06}, R>4R_{\oplus} 
\end{equation} 

Again using the same method as \citet{Lissauer2011b}, we measured the mutual Hill radii for the six consecutive planet pairs. When the following equation holds true, the pair of planets are Hill stable, meaning that they will never develop crossing orbits, assuming circular, coplanar orbits.  Whether the equation holds true depends more heavily on stellar radius (via the $R_{\rm{PL}}/R_{*}$ and the mass-radius relationship) than stellar mass (via $a/R_{*}$ and Newton's Third Law).  The best fit values for these two parameters from Y$^2$ interpolation are $R=1.04^{+0.12}_{-0.10}R_{\odot}$ and $M=0.99^{+0.10}_{-0.10}M_{\odot}$, consistent within $1\sigma$ of the stellar parameters from \citet{Huber2014}.  The Hill stability metric, where $a$ is the semi-major axis of the planet and $R_H$ is the Hill radius, is:  

\begin{equation}
\Delta = \frac{a_{\rm{outer}} - a_{\rm{inner}}}{R_{H}} > 2\sqrt{3} 	\approx 3.46
\end{equation} 

Because of the extremely large TTVs, we might expect that the mutual Hill spheres of the outer two planets are very close to the Hill stability criterion.  As expected from the TTVs, the least Hill stable pair of planets are KOI-351.01 ($P=331.6$ days) and KOI-351.02 ($P=210.6$ days), the outer two planets, with $\Delta=5.60$.  All planet pairs remain Hill stable across the $3\sigma$ range of stellar radius and mass. 

Lastly, we ran an orbital integration with the \textit{Mercury} code \citep{Chambers1999} using the best fit values and for planetary masses corresponding to $\pm1\sigma$ in stellar mass and radius, assuming coplanarity and circular orbits. The systems were stable for $>100$ Myr.  However, doubling the mass of the outer planet resulted in instability, suggesting that the mass of the outer planet is on the less massive end of the flat part of the mass-radius relationship for gas giants.  These orbital integrations are only feasibility tests and are by no means an exhaustive analysis proving dynamic stability.  

\begin{figure*}[tp]
	\centering
		\centerline{\includegraphics{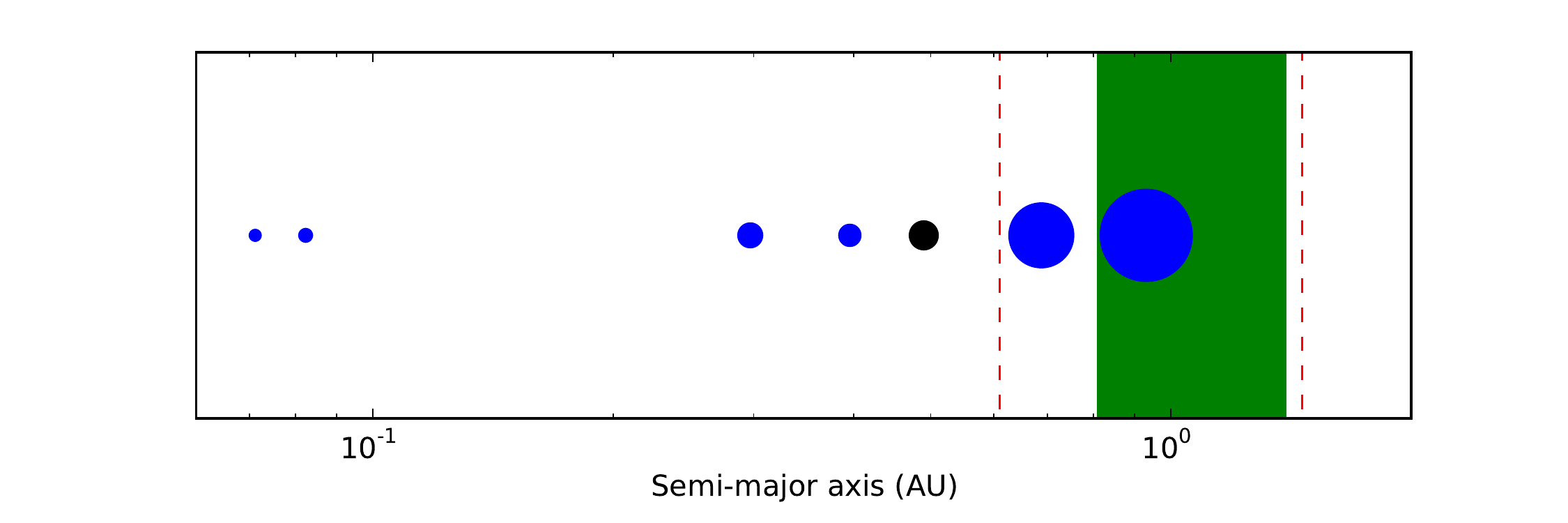}}
	\caption{Orbital representation of KOI-351's seven planet candidates with the habitable zone as reference, the black circle being the new candidate.  The relative planetary radii are shown as the size of the symbol.}
	\label{fig:koi351}
\end{figure*}

KOI-351 is also interesting as it is close to being a compact analog to the solar system; see Figure~\ref{fig:koi351} for an orbital representation of the system.  The nominal radius values of the inner five planet candidates are all in the Earth to mini-Neptune regime ($\lesssim3.0R_{\oplus}$), while the outer two candidates are gas giants.  This differs from Kepler-11 \citep{Lissauer2011a}, in that all of Kepler-11's planets are in the super-Earth to Neptune sizes.  However, we stress that the error bars on the KOI-351's stellar radius and thus the planetary radii are very large at roughly 50\% each.  The new candidate has a radius of $R=2.70\pm2.07	R_{\oplus}$.  KOI~351 deserves a strong follow-up observation to better constrain the radii, analyze the TTVs, and confirm these planets.  This new candidate also exemplifies how complicated signals that may confuse or overload computers can be deciphered by visual checks.  We also note that Planet Hunters had independently discovered the 8.72 and the 91.94 day signals. However, during preparation of this paper, both were upgraded to KOI candidate status. 

\section{Conclusions}
\label{conc}

Planet Hunters is designed to be complementary to the \textit{Kepler} team's own planet search algorithm, TPS, which has proven itself to be extremely successful with over 18,000 TCEs and more than 3,500 candidates discovered. One of the most important discoveries presented here is the addition of one new planetary candidate around KOI-351, a known six planet candidate system.  According to \texttt{http://exoplanets.org/} \citep{Wright2011}, the two stars with the largest number of confirmed planets, excluding our Sun, contain six planets each:  HD~10180 \citep{Lovis2011} and Kepler-11 \citep{Lissauer2011a}. KOI-435 is the only other \textit{Kepler} system with six planet candidates \cite{Ofir2013}.  Furthermore, there are currently only five exoplanetary systems with five confirmed planets \citep{Wright2011}: 55 Cancri \citep{Fischer2008}, Kepler-20 \citep{Borucki2011,Fressin2012}, Kepler-32 \citep{Borucki2011,Swift2013}, Kepler-33 \citep{Borucki2011,Lissauer2012}, and Kepler-62 \citep{Borucki2013}.  Although HD~10180 has been claimed to have seven \citep{Lovis2011} to nine \citep{Tuomi2012} periodic signals, and GJ 667C is claimed to have up to seven periodic signals \citep{Anglada2013}, the planetary nature of those signals is yet to be confirmed.  Conversely, we believe that KOI-351 is a true seven planet system with the highest level of certainty short of official confirmation.  Analysis shows that this seventh signal is not an alias of the other six and is feasibly stable.  Large TTVs for the outer two planet pairs strongly indicate that the outer two planet pairs are likely interacting gravitationally, which helps to validate the system.  It is also well known that candidates in multiple candidate systems have much lower false positive probabilities than single candidate systems \citep{Lissauer2012}.

KOI-351's system looks somewhat like our own, but much more compact; all seven planet candidates are $\lesssim1$ AU from their host star. While the radii have large error bars, all five of the inner planets have sub-Neptune radii, with KOI-351.07 being the outermost of these.  The outer two planets appear to be gas giants.  The high multiplicity of this system may merit an increased scrutiny of known planetary systems for additional planet candidates.  

The most important contribution Planet Hunters provides are the considerable number of new long period candidates.  These long period planets can probe the HZ of solar-like stars.  Indeed half of the 14 planet candidates reported here are located in their host star's HZ.  One such candidate (KIC 11152511) even straddles the transition between super Earth and mini-Neptune radii, making it especially deserving of follow-up analysis.   However, many of these new candidates contain only two transits.  With the failure of \textit{Kepler}'s third reaction wheel, one and two transit systems become important in order to study cool planets orbiting far from their host stars.  This is especially so for known planetary systems, since the probability that a false positive occurs in a known planetary system is much lower \citep{Lissauer2012,Lissauer2014}.  Planet Hunters continue to collect one and two transit systems, and these will be further explored in a future paper, Picard et al. (2014, in preparation).

Also noteworthy is the new planet candidate around KIC~2437209, which boasts a 74 hour transit every 281 days.  Two scenarios fit the transit light curve:  an evolved star with a large secondary object transiting the star and a highly eccentric gas giant.  If this is indeed a highly eccentric $e=0.974$ gas giant, as the long duration might suggest, it may be a planet potentially undergoing high eccentricity migration.  However, its position in the $g'$ vs. $(g'-r')$ color-magnitude diagram places KIC~2437209 on the giant branch of NGC~6791.  

This paper brings the total new planetary candidates discovered by Planet Hunters to $\sim60$ plus the two confirmed planets, PH1~b and PH2~b, and new candidates continue to be passed to the science team.  Planet Hunters will continue their search for more planetary candidates in the archived \textit{Kepler} data.  The failure of \textit{Kepler}'s third reaction wheel has led to a proposal for a two wheeled extended mission, K2, which plans to observe new fields of view in the ecliptic for 75 days each \citep{Howell2014}.  Should the K2 \textit{Kepler} mission be approved, Planet Hunters will be analyzing each new campaign as well.  

\noindent{\it Acknowledgements} 

We thank all Planet Hunter volunteers, who were indispensable for their work in discovering and analyzing planet candidates in this paper.  All Planet Hunters are individually acknowledged at \url{http://www.planethunters.org/authors}.  The authors thank the Planet Hunters volunteers who participated in identifying and analyzing the candidates presented in this paper. They are individually recognized at \url{http://www.planethunters.org/PH6}.

We thank the anonymous referee for helpful comments and for suggesting showcasing Planet Hunters considerable contributions to the number of long period candidates.

DF acknowledges funding support for PlanetHunters.org from Yale University and support from the NASA Supplemental Outreach Award, 10-OUTRCH.210-0001 and the NASA ADAP12-0172.  TSB acknowledges support provided through NASA grant ADAP12-0172.  KS gratefully acknowledges support from Swiss National Science Foundation Grant PP00P2\_138979/1.  Planet Hunters is partially supported by NASA JPL's PlanetQuest program. The Talk system was supported by the National Science Foundation under Grant No. DRL-0941610.   The Zooniverse Project is supported by The Leverhulme Trust and by the Alfred P. Sloan foundation.   We have used public data from NASA/IPAC/NExScI Star and the Exoplanet Database, which is maintained by JPL at Caltech, under contract with NASA. Our research has utilized NASA's Astrophysics Data System Bibliographic Services.  We acknowledge the \textit{Kepler} science team and others involved for their great and ongoing work.  Funding for the \textit{Kepler} mission is provided by the NASA Science Mission directorate. The publicly released \textit{Kepler} light curves were downloaded from the Multimission Archive at the Space Telescope Science Institute (MAST). STScI is operated by the Association of Universities for Research in Astronomy, Inc., under NASA contract NAS5-26555. Support for MAST for non-HST data is provided by the NASA Office of Space Science via grant NNX09AF08G and by other grants and contracts.  This work made use of PyKE, a software package for the reduction and analysis of \textit{Kepler} data. This open source software project is developed and distributed by the NASA \textit{Kepler} Guest Observer Office. This research has made use of the NASA Exoplanet Archive, which is operated by the California Institute of Technology, under contract with the National Aeronautics and Space Administration under the Exoplanet Exploration Program.




\clearpage


\begin{deluxetable}{lcccc}
\tablewidth{0pt}
\tablecaption{Visual companion detections with UKIRT images.\label{tab:AO}}
\tablehead{
\colhead{\textbf{KIC}} &
\colhead{\textbf{$\Delta$ K$_P$}} &
\colhead{\textbf{Separation}} &
\colhead{\textbf{Significance}} &
\colhead{\textbf{PA}} \\
\colhead{\textbf{}} &
\colhead{\textbf{(mag)}} &
\colhead{\textbf{(arcsec)}} &
\colhead{\textbf{($\sigma$)}} &
\colhead{\textbf{(deg)}} \\ }

\startdata

2437209   & \nodata  & \nodata  & \nodata  & \nodata  \\
5010054   & \nodata  & \nodata  & \nodata  & \nodata  \\
5094412   & \nodata  & \nodata  & \nodata  & \nodata  \\
5522786   & \nodata  & \nodata  & \nodata  & \nodata  \\
5732155   & \nodata  & \nodata  & \nodata  & \nodata  \\
6372194   & 2.0      & 1.7      & \nodata  & 277.6    \\
6436029   & \nodata  & \nodata  & \nodata  & \nodata  \\
6805414   & \nodata  & \nodata  & \nodata  & \nodata  \\
9662267   & \nodata  & \nodata  & \nodata  & \nodata  \\
9704149   & \nodata  & \nodata  & \nodata  & \nodata  \\
10255705  & 2.1      & 1.4      & \nodata  & 165.0    \\
11152511  & \nodata  & \nodata  & \nodata  & \nodata  \\
11442793  & \nodata  & \nodata  & \nodata  & \nodata  \\
12454613  & \nodata  & \nodata  & \nodata  & \nodata  \\
 
\enddata

\tablecomments{Nearest neighbors of the planet candidate hosting stars out to 4\arcsec.  Only KIC 6805414 has a companion significantly above the background.  KIC 10255705 has a neighbor within its $3\sigma$ confusion radius.}

\end{deluxetable}

\begin{landscape}

\begin{deluxetable}{lccccccc|ccccccc}
\tabletypesize{\tiny}

\tablewidth{0pt}
\tablecaption{Stellar Parameters\label{tab:st}}

\tablehead{
 \multicolumn{8}{c}{\textbf{Input Values}}  &  \multicolumn{7}{c}{\textbf{Derived Values}} \\ 
\colhead{\textbf{Star}} &
\colhead{\textbf{$Kp$}} &
\colhead{\textbf{($g-r$)}} &
\colhead{\textbf{$T_{\rm eff}$}} &
\colhead{\textbf{$\log g$}} 	&
\colhead{\textbf{M$_{\ast}$}} &
\colhead{\textbf{R$_{\ast}$}}	&
\colhead{\textbf{[Fe/H]}} &
\colhead{\textbf{$T_{\rm eff}$}} &
\colhead{\textbf{$\log g$}} 	&
\colhead{\textbf{M$_{\ast}$}} &
\colhead{\textbf{R$_{\ast}$}}	&
\colhead{\textbf{[Fe/H]}} &
\colhead{\textbf{L$_{\ast}$}}	&
\colhead{\textbf{$\rho_{\ast}$}} \\
\colhead{\textbf{(KIC)}} &
\colhead{\textbf{(mag)}}	&
\colhead{\textbf{(mag)}}	&
\colhead{\textbf{(K)}}	&
\colhead{\textbf{(cgs)}}	&
\colhead{\textbf{(M$_{\odot}$)}} &
\colhead{\textbf{(R$_{\odot}$)}} &
\colhead{\textbf{(dex)}}	&
\colhead{\textbf{(K)}}	&
\colhead{\textbf{(cgs)}}	&
\colhead{\textbf{(M$_{\odot}$)}} &
\colhead{\textbf{(R$_{\odot}$)}} &
\colhead{\textbf{(dex)}}	&
\colhead{\textbf{(L$_{\odot}$)}} &
\colhead{\textbf{(g~cm$^{-3}$)}} \\ 
}

\startdata

2437209 & 16.353 & 1.001 & 4822 & 4.56 & 0.73 & 0.74 & -0.04 & $4842^{+157}_{-90}$ & $3.07^{+0.60}_{-0.15}$ & $1.16^{+0.75}_{-0.30}$ & $4.97^{+2.09}_{-1.14}$ & $0.07^{+0.45}_{-0.23}$ & $10.68^{+14.48}_{-5.52}$ & $2.33\pm0.49$ \\ 
5010054 & 13.961 & 0.631 & 5944 & 4.22 & 1.05 & 1.32 & 0.02 & $5948^{+145}_{-81}$ & $4.37^{+0.15}_{-0.15}$ & $1.07^{+0.11}_{-0.11}$ & $1.05^{+0.38}_{-0.21}$ & $0.01^{+0.20}_{-0.08}$ & $1.17^{+1.15}_{-0.12}$ & $1.30\pm0.66$ \\ 
5094412 & 15.772 & 0.633 & 5511 & 4.60 & 0.83 & 0.76 & -0.34 & $5572^{+109}_{-96}$ & $4.58^{+0.15}_{-0.15}$ & $0.82^{+0.08}_{-0.08}$ & $0.76^{+0.15}_{-0.15}$ & $-0.42^{+0.20}_{-0.11}$ & $0.46^{+0.07}_{-0.05}$ & $2.64\pm0.70$ \\ 
5522786\tablenotemark{a} & 9.350 & -0.146 & 8941 & 4.12 & 2.10 & 2.08 & 0.07 & \nodata & \nodata & \nodata & \nodata & \nodata & $24.78^{+17.88}_{-17.88}$ & $0.328^{+0.234}_{-0.223}$ \\
5732155 & 15.195 & 0.550 & 6140 & 4.45 & 1.09 & 1.03 & -0.04 & $6065^{+131}_{-77}$ & $4.41^{+0.15}_{-0.15}$ & $1.02^{+0.10}_{-0.21}$ & $1.06^{+0.21}_{-0.11}$ & $-0.12^{+0.15}_{-0.10}$ & $1.17^{+0.31}_{-0.12}$ & $1.37\pm0.36$ \\ 
6372194 & 15.870 & 0.698 & 5314 & 4.62 & 0.80 & 0.72 & -0.34 & $5308^{+187}_{-68}$ & $4.62^{+0.15}_{-0.15}$ & $0.79^{+0.08}_{-0.08}$ & $0.71^{+0.14}_{-0.14}$ & $-0.44^{+0.22}_{-0.09}$ & $0.35^{+0.07}_{-0.04}$ & $3.17\pm0.73$ \\ 
6436029 & 15.768 & 1.007 & 4817 & 4.50 & 0.80 & 0.83 & 0.42 & $4878^{+173}_{-31}$ & $4.55^{+0.15}_{-0.15}$ & $0.83^{+0.08}_{-0.08}$ & $0.81^{+0.16}_{-0.16}$ & $0.42^{+0.05}_{-0.03}$ & $0.36^{+0.09}_{-0.04}$ & $2.20\pm0.69$ \\ 
6805414 & 15.392 & 0.555 & 6108 & 4.47 & 1.07 & 1.00 & -0.08 & $6029^{+133}_{-68}$ & $4.43^{+0.15}_{-0.15}$ & $1.00^{+0.10}_{-0.10}$ & $1.02^{+0.20}_{-0.20}$ & $-0.25^{+0.20}_{-0.06}$ & $1.20^{+0.20}_{-0.12}$ & $1.36\pm0.60$ \\ 
9662267 & 14.872 & 0.534 & 6032 & 4.49 & 1.06 & 0.97 & -0.06 & $5870^{+172}_{-43}$ & $4.44^{+0.15}_{-0.15}$ & $1.02^{+0.10}_{-0.10}$ & $0.98^{+0.20}_{-0.20}$ & $-0.21^{+0.17}_{-0.08}$ & $1.02^{+0.20}_{-0.10}$ & $1.56\pm0.65$ \\ 
9704149 & 15.102 & 0.540 & 5897 & 4.53 & 0.98 & 0.89 & -0.16 & $5738^{+165}_{-47}$ & $4.50^{+0.15}_{-0.15}$ & $0.91^{+0.09}_{-0.09}$ & $0.88^{+0.18}_{-0.18}$ & $-0.37^{+0.18}_{-0.07}$ & $0.79^{+0.11}_{-0.08}$ & $1.89\pm0.39$ \\ 
10255705 & 12.950 & 0.698 & 5286 & 3.96 & 0.90 & 1.64 & -0.12 & $5339^{+153}_{-84}$ & $3.73^{+0.80}_{-0.15}$ & $1.34^{+0.31}_{-0.26}$ & $2.53^{+1.13}_{-0.58}$ & $-0.20^{+0.35}_{-0.10}$ & $4.29^{+4.76}_{-1.71}$ & $0.11\pm0.10$ \\ 
11152511 & 13.618 & 0.640 & 5568 & 4.27 & 0.86 & 1.12 & -0.08 & $5548^{+161}_{-58}$ & $4.44^{+0.15}_{-0.15}$ & $0.92^{+0.10}_{-0.09}$ & $0.92^{+0.25}_{-0.18}$ & $-0.08^{+0.28}_{-0.09}$ & $0.64^{+0.46}_{-0.06}$ & $1.53\pm0.99$ \\ 
11442793 & 13.804 & 0.398 & 6238 & 4.38 & 0.97 & 1.05 & -0.40 & $6258^{+150}_{-103}$ & $4.39^{+0.15}_{-0.15}$ & $0.99^{+0.10}_{-0.10}$ & $1.04^{+0.21}_{-0.21}$ & $-0.48^{+0.23}_{-0.09}$ & $1.47^{+0.48}_{-0.18}$ & $1.24\pm0.43$ \\ 
12454613 & 13.537 & 0.617 & 5530 & 4.56 & 0.94 & 0.84 & 0.00 & $5436^{+144}_{-54}$ & $4.55^{+0.15}_{-0.15}$ & $0.87^{+0.09}_{-0.09}$ & $0.82^{+0.16}_{-0.16}$ & $-0.17^{+0.18}_{-0.08}$ & $0.55^{+0.06}_{-0.06}$ & $2.34\pm0.69$ \\

\enddata
\tablenotetext{a}{KIC 5522786 has a $T_{\rm{eff}}$ above the limits of our stellar modeling, so we used the stellar parameters from \citep{Huber2014}.}
\tablecomments{Stellar inputs \citep{Huber2014}  and outputs from the iterative light curve and stellar isochrone fitting routine. See Section~\ref{sec:stellarchar} for details.}

\end{deluxetable}

\end{landscape}

\begin{landscape}
\begin{deluxetable}{lccccccccccccc}
\tabletypesize{\tiny}
\tablewidth{0pt}
\tablecaption{Orbital Parameters\label{tab:pl}}
\tablehead{
\colhead{\textbf{Star}} &
\colhead{\textbf{$T_0$}} &
\colhead{\textbf{Period}} &
\colhead{\textbf{Impact}} &
\colhead{\textbf{R$_{\rm PL}$/R$_{\ast}$}} &
\colhead{\textbf{e}} &
\colhead{\textbf{$\omega$}} &
\colhead{\textbf{$i$}} &
\colhead{\textbf{a/R$_{\ast}$}}	&
\colhead{\textbf{a}}	&
\colhead{\textbf{R$_{\rm PL}$}} 	&
\colhead{\textbf{S}}	&
\colhead{\textbf{Depth}}	&
\colhead{\textbf{Duration}}  \\
\colhead{\textbf{(KIC)}} &
\colhead{\textbf{(MJD)}}	&
\colhead{\textbf{(d)}}	&
\colhead{\textbf{Parameter}}	&
\colhead{\textbf{ }}	&
\colhead{\textbf{ }}	&
\colhead{\textbf{(radian)}}	&
\colhead{\textbf{($\deg$)}} &
\colhead{\textbf{}} &
\colhead{\textbf{(AU)}} &
\colhead{\textbf{(R$_{\oplus}$)}}	&
\colhead{\textbf{($S_0$)}} &
\colhead{\textbf{(ppm)}} & 
\colhead{\textbf{(hours)}} 
}

\startdata

2437209 & 55329.2790 & $281.3290^{+0.0012}_{-0.0040}$ & $0.48^{+0.10}_{-0.48}$ & $0.1154^{+0.0006}_{-0.0017}$ & $0.43^{+0.23}_{-0.43}$ & $4.72^{+0.50}_{-2.38}$ & $89.31^{+0.24}_{-1.26}$ & $39.79^{+25.10}_{-7.25}$ & $0.88\pm0.13$ & $62.64\pm19.61$ & $14.39\pm11.76$ & $13115\pm1137$ & 73.56 \\ 
5010054\tablenotemark{a} & 55188.9590 & $904.0905^{+0.0114}_{-0.0533}$ & $0.03^{+0.84}_{-0.03}$ & $0.0257^{+0.0006}_{-0.0010}$ & $0.31^{+0.68}_{-0.31}$ & $3.72^{+1.43}_{-3.72}$ & $90.00^{+0.00}_{-0.31}$ & $372.31^{+38.58}_{-8.93}$ & $1.86\pm0.09$ & $2.92\pm0.48$ & $0.37\pm0.07$ & $789\pm136$ & 21.23 \\ 
5094412 & 55009.0210 & $276.8800^{+0.0000}_{-0.0000}$ & $0.99^{+0.01}_{-0.09}$ & $0.0895^{+0.0187}_{-0.0591}$ & $0.04^{+0.10}_{-0.04}$ & $5.97^{+0.32}_{-5.97}$ & $89.74^{+0.21}_{-0.17}$ & $220.60^{+12.59}_{-13.51}$ & $0.77\pm0.03$ & $7.38\pm3.24$ & $0.77\pm0.17$ & $2014\pm280$ & 3.77 \\ 
5522786\tablenotemark{a} & 55115.4928 & $757.1570^{+0.0089}_{-0.0299}$ & $0.36^{+0.40}_{-0.29}$ & $0.0089^{+0.0003}_{-0.0014}$ & $0.45^{+0.41}_{-0.45}$ & $1.53^{+2.79}_{-1.53}$ & $89.91^{+0.09}_{-0.28}$ & $236.70^{+25.98}_{-20.35}$ & $2.07\pm0.06$ & $1.86\pm0.25$ & $4.85\pm0.98$ & $90\pm13$ & 14.71 \\ 
5732155\tablenotemark{a} & 55369.2000 & $644.1978^{+0.0077}_{-0.0182}$ & $0.39^{+0.12}_{-0.39}$ & $0.0573^{+0.0021}_{-0.0028}$ & $0.52^{+0.20}_{-0.24}$ & $4.40^{+0.34}_{-0.52}$ & $89.93^{+0.03}_{-0.12}$ & $307.37^{+22.70}_{-13.83}$ & $1.47\pm0.09$ & $6.34\pm0.48$ & $0.59\pm0.10$ & $3644\pm398$ & 24.03 \\ 
6372194 & 55375.9140 & $281.5904^{+0.0031}_{-0.0094}$ & $0.97^{+0.03}_{-0.11}$ & $0.1027^{+0.1905}_{-0.0086}$ & $0.27^{+0.09}_{-0.27}$ & $2.34^{+3.95}_{-2.34}$ & $89.76^{+0.01}_{-0.07}$ & $232.59^{+12.76}_{-10.00}$ & $0.78\pm0.02$ & $7.90\pm7.68$ & $0.59\pm0.18$ & $10515\pm344$ & 5.90 \\ 
6436029 & 55290.6000 & $505.4611^{+0.0152}_{-0.0340}$ & $0.56^{+0.43}_{-0.56}$ & $0.0470^{+0.0680}_{-0.0228}$ & $0.63^{+0.36}_{-0.63}$ & $4.63^{+1.66}_{-4.63}$ & $89.90^{+0.08}_{-0.08}$ & $308.28^{+11.00}_{-29.10}$ & $1.17\pm0.04$ & $4.16\pm4.04$ & $0.26\pm0.07$ & $1972\pm389$ & 13.49 \\ 
6805414 & 55137.9480 & $200.2473^{+0.0010}_{-0.0027}$ & $0.39^{+0.12}_{-0.02}$ & $0.1083^{+0.0001}_{-0.0006}$ & $0.74^{+0.09}_{-0.05}$ & $4.86^{+0.27}_{-0.57}$ & $89.84^{+0.02}_{-0.03}$ & $142.17^{+12.84}_{-8.93}$ & $0.67\pm0.04$ & $11.91\pm1.58$ & $2.67\pm0.50$ & $12482\pm296$ & 23.72 \\ 
9662267\tablenotemark{a} & 55314.3800 & $466.1710^{+0.0112}_{-0.0370}$ & $0.74^{+0.19}_{-0.26}$ & $0.0426^{+0.0052}_{-0.0094}$ & $0.65^{+0.34}_{-0.65}$ & $5.37^{+0.92}_{-5.37}$ & $89.83^{+0.02}_{-0.11}$ & $257.93^{+22.65}_{-16.43}$ & $1.18\pm0.04$ & $4.50\pm0.98$ & $0.77\pm0.15$ & $1511\pm229$ & 10.55 \\ 
9704149\tablenotemark{a} & 55252.2220 & $697.0159^{+0.0000}_{-0.0000}$ & $0.38^{+0.34}_{-0.38}$ & $0.0524^{+0.0039}_{-0.0018}$ & $0.03^{+0.11}_{-0.03}$ & $6.29^{+0.00}_{-6.29}$ & $89.94^{+0.02}_{-0.05}$ & $367.03^{+23.65}_{-10.04}$ & $1.49\pm0.06$ & $5.01\pm0.39$ & $0.33\pm0.09$ & $3214\pm406$ & 13.86 \\ 
10255705\tablenotemark{a} & 55378.2200 & $707.8201^{+0.0000}_{-0.0000}$ & $0.57^{+0.09}_{-0.57}$ & $0.0362^{+0.0010}_{-0.0120}$ & $0.52^{+0.18}_{-0.52}$ & $4.82^{+1.47}_{-4.82}$ & $89.76^{+0.07}_{-0.19}$ & $136.78^{+55.92}_{-6.74}$ & $1.69\pm0.13$ & $10.08\pm3.09$ & $1.67\pm0.66$ & $1118\pm102$ & 39.72 \\ 
11152511 & 55193.2608 & $287.3377^{+0.0348}_{-0.0837}$ & $0.36^{+0.49}_{-0.36}$ & $0.0187^{+0.0113}_{-0.0166}$ & $0.00^{+0.08}_{-0.00}$ & $0.00^{+6.16}_{-0.00}$ & $89.89^{+0.11}_{-4.14}$ & $190.60^{+4.72}_{-48.07}$ & $0.83\pm0.05$ & $1.93\pm1.50$ & $1.06\pm0.80$ & $463\pm95$ & 9.38 \\ 
11442793 & 55087.1710 & $124.9199^{+0.0236}_{-0.0540}$ & $0.54^{+0.05}_{-0.43}$ & $0.0239^{+0.0147}_{-0.0215}$ & $0.69^{+0.30}_{-0.69}$ & $4.73^{+0.25}_{-1.22}$ & $89.69^{+0.31}_{-3.86}$ & $99.70^{+10.81}_{-4.08}$ & $0.49\pm0.03$ & $2.70\pm2.07$ & $5.82\pm1.70$ & $480\pm87$ & 10.48 \\ 
12454613\tablenotemark{a} & 55322.7620 & $736.3819^{+0.0065}_{-0.0211}$ & $0.37^{+0.42}_{-0.37}$ & $0.0289^{+0.0064}_{-0.0015}$ & $0.16^{+0.21}_{-0.16}$ & $0.00^{+0.17}_{-0.00}$ & $89.95^{+0.03}_{-0.18}$ & $399.19^{+22.45}_{-17.10}$ & $1.54\pm0.04$ & $2.56\pm0.42$ & $0.23\pm0.04$ & $1006\pm142$ & 12.79 \\ 

\enddata

\tablenotetext{a}{Planet candidates with only two transits.  Of these, KIC 9704149 actually only has 1.5 transits due to a data gap interrupting the second transit.}
\tablecomments{Orbital fit to the \textit{Kepler} light curves from the iterative light curve and stellar isochrone fitting routine.}

\end{deluxetable}

\begin{landscape}
\begin{deluxetable}{lccccccccccc}
\tabletypesize{\tiny}
\tablewidth{0pt}
\tablecaption{KOI-351 Candidates\label{tab:351}}
\tablehead{
\colhead{\textbf{Candidate}}          &
\colhead{\textbf{$T_0$}}      &
\colhead{\textbf{Period}}      & 
\colhead{\textbf{Impact}}      &
\colhead{\textbf{R$_{\rm PL}$/R$_{\ast}$}} &
\colhead{\textbf{$i$}} &
\colhead{\textbf{a/R$_{\ast}$}}	&
\colhead{\textbf{a}}	&
\colhead{\textbf{R$_{\rm PL}$}} 	&
\colhead{\textbf{S}}	&
\colhead{\textbf{Depth}}	&
\colhead{\textbf{Duration}}  \\
\colhead{\textbf{}} &
\colhead{\textbf{(MJD)}}	&
\colhead{\textbf{(d)}}	&
\colhead{\textbf{Parameter}}	&
\colhead{\textbf{ }}	&
\colhead{\textbf{($\deg$)}} &
\colhead{\textbf{}} &
\colhead{\textbf{(AU)}} &
\colhead{\textbf{(R$_{\oplus}$)}}	&
\colhead{\textbf{($S_0$)}} &
\colhead{\textbf{(ppm)}} & 
\colhead{\textbf{(hours)}} \\
}
\startdata

KOI-351.01  & 54972.98  & 331.64  & 0.326  & 0.0833  & 89.95  & 180    & 0.94   & $9.4\pm4.0$   & 1.67   & 8277  & 14.54  \\ 
KOI-351.02  & 54979.55  & 210.60  & 0.313  & 0.0584  & 89.95  & 133    & 0.69   & $6.6\pm2.9$   & 3.09   & 4073  & 12.23  \\ 
KOI-351.03  & 54991.45  & 59.74   & 0.01   & 0.0217  & 89.95  & 57.4   & 0.30   & $2.5\pm1.1$   & 16.3   & 567   & 8.12   \\ 
KOI-351.04  & 54966.82  & 91.94   & 0.29   & 0.0192  & 89.95  & 76.5   & 0.40   & $2.2\pm1.0$   & 9.12   & 432   & 8.97   \\ 
KOI-351.05  & 54971.96  & 8.72    & 0.36   & 0.0116  & 88.81  & 15.9   & 0.083  & $1.3\pm0.51$  & 213    & 148   & 3.97   \\ 
KOI-351.06  & 54970.17  & 7.01    & 0.34   & 0.0099  & 88.81  & 13.8   & 0.072	& $1.1\pm0.48$  & 284    & 104   & 3.71   \\ 
KOI-351.07  & 55087.22  & 124.92  & 0.54   & 0.0239  & 89.69  & 99.8   & 0.49   & $2.7\pm2.07$  & 6.12   & 480   & 10.48  \\ 

\enddata

\tablecomments{KOI-351.01 through KOI-351.06 values are directly from the public KOI data (http://exoplanetarchive.ipac.caltech.edu, last accessed March 11, 2014), scaled to our model's stellar parameters of $M=0.99\pm0.10 M_{\odot}$, $R=1.04^{+0.12}_{-0.10}R_{\odot}$, and $T_{\rm{eff}}=6258^{+150}_{−103}$ K.}

\end{deluxetable}
\end{landscape}

\end{landscape}

\begin{landscape}
\begin{deluxetable}{lcccccccccccccc}
\tabletypesize{\tiny}
\tablewidth{0pt}
\tablecaption{Nearest linear ephemerides of KOI-351's previously known candidates to KOI-351.07\label{tab:ephem}}
\tablehead{
\colhead{\textbf{ }}         &
\colhead{\textbf{T0}}        & 
\colhead{\textbf{Phase}}     & 
\colhead{\textbf{T2}}        & 
\colhead{\textbf{Phase}}     & 
\colhead{\textbf{T5}}        & 
\colhead{\textbf{Phase}}     & 
\colhead{\textbf{T6}}        & 
\colhead{\textbf{Phase}}     & 
\colhead{\textbf{T8}}        & 
\colhead{\textbf{Phase}}     & 
\colhead{\textbf{T9}}        & 	
\colhead{\textbf{Phase}}     & 
\colhead{\textbf{Depth}}     &  
\colhead{\textbf{Duration}}  \\
\colhead{\textbf{ }}                   &
\colhead{\textbf{254.64}}              & 
\colhead{\textbf{Difference}}          & 
\colhead{\textbf{504.57$^{\dagger}$}}  & 
\colhead{\textbf{Difference}}          &
\colhead{\textbf{879.27}}              & 
\colhead{\textbf{Difference}}          &
\colhead{\textbf{1004.17}}             & 
\colhead{\textbf{Difference}}          &
\colhead{\textbf{1253.97}}             & 
\colhead{\textbf{Difference}}          &
\colhead{\textbf{1378.87}}             &  
\colhead{\textbf{Difference}}          &
\colhead{\textbf{(ppm)}}               &
\colhead{\textbf{(hours)}}             \\
}
\startdata
KOI-351.01  & 140.48  & \nodata  & 472.12  & \nodata  & 803.76   & \nodata  & 803.76   & \nodata  & 1135.41  & \nodata  & 1135.41  & \nodata  & 8423     & 14.53    \\
KOI-357.07  & 254.64  & 0.34     & 504.57  & 0.10     & 879.27   & 0.23     & 1004.17  & 0.60     & 1253.97  & 0.36     & 1378.87  & 0.73     & 433      & 10.75    \\
KOI-351.01  & 472.12  & \nodata  & 803.76  & \nodata  & 1135.41  & \nodata  & 1135.41  & \nodata  & 1467.41  & \nodata  & 1467.41  & \nodata  & \nodata  & \nodata  \\ \hline
KOI-351.02  & 147.05  & \nodata  & 357.66  & \nodata  & 778.87   & \nodata  & 989.47   & \nodata  & 1200.08  & \nodata  & 1200.08  & \nodata  & 4150     & 11.69    \\
KOI-357.07  & 254.64  & 0.51     & 504.57  & 0.70     & 879.27   & 0.48     & 1004.17  & 0.07     & 1253.97  & 0.26     & 1378.87  & 0.85     & 433      & 10.75    \\
KOI-351.02  & 357.66  & \nodata  & 568.26  & \nodata  & 989.47	 & \nodata  & 1200.08  & \nodata  & 1410.68  & \nodata  & 1410.68  & \nodata  & \nodata  & \nodata  \\ \hline
KOI-351.03  & 218.69  & \nodata  & 457.65  & \nodata  & 875.83   & \nodata  & 995.31   & \nodata  & 1234.27  & \nodata  & 1353.75  & \nodata  & 587      & 8.07     \\
KOI-357.07  & 254.64  & 0.60     & 504.57  & 0.79     & 879.27   & 0.06     & 1004.17  & 0.15     & 1253.97  & 0.33     & 1378.87  & 0.42     & 433      & 10.75    \\
KOI-351.03  & 278.43  & \nodata  & 517.39  & \nodata  & 935.57   & \nodata  & 1055.05  & \nodata  & 1294.01  & \nodata  & 1413.49  & \nodata  & \nodata  & \nodata  \\ \hline
KOI-351.04  & 226.25  & \nodata  & 502.06  & \nodata  & 869.80   & \nodata  & 961.74   & \nodata  & 1237.55  & \nodata  & 1329.48  & \nodata  & 410      & 6.42     \\ 
KOI-357.07  & 254.64  & 0.31     & 504.57  & 0.03     & 879.27   & 0.10     & 1004.17  & 0.46     & 1253.97  & 0.18     & 1378.87  & 0.54     & 433      & 10.75    \\
KOI-351.04  & 318.19  & \nodata  & 594.00  & \nodata  & 961.74   & \nodata  & 1053.68  & \nodata  & 1329.48  & \nodata  & 1421.42  & \nodata  & \nodata  & \nodata  \\ \hline
KOI-351.05  & 252.84  & \nodata  & 497.00  & \nodata  & 871.96   & \nodata  & 1002.76  & \nodata  & 1246.92  & \nodata  & 1377.72  & \nodata  & 116      & 3.69     \\
KOI-357.07  & 254.64  & 0.21     & 504.57  & 0.87     & 879.27   & 0.84     & 1004.17  & 0.16     & 1253.97  & 0.81     & 1378.87  & 0.13     & 433      & 10.75    \\
KOI-351.05  & 261.56  & \nodata  & 505.72  & \nodata  & 880.68   & \nodata  & 1011.48  & \nodata  & 1255.64  & \nodata  & 1386.44  & \nodata  & \nodata  & \nodata  \\ \hline
KOI-351.06  & 249.80  & \nodata  & 502.10  & \nodata  & 873.54   & \nodata  & 999.69   & \nodata  & 1251.99  & \nodata  & 1378.14  & \nodata  & 91       & 3.80     \\ 
KOI-357.07  & 254.64  & 0.69     & 504.57  & 0.35     & 879.27   & 0.82     & 1004.17  & 0.64     & 1253.97  & 0.28     & 1378.87  & 0.10     & 433      & 10.75    \\
KOI-351.06  & 256.81  & \nodata  & 509.11  & \nodata  & 880.55   & \nodata  & 1006.70  & \nodata  & 1259.00  & \nodata  & 1385.15  & \nodata  & \nodata  & \nodata  \\
\enddata

\tablecomments{$^{\dagger}$ Transit 5 (T5) is only a partial transit.  A data gap within quarter 5 blocks out everything but the egress.  \\
Midtransit times of our new KOI-351.07 sandwiched between the nearest two midtransit times of all other six candidates in the system, according to a linear ephemeris as calculated by the NASA Exoplanet Archive's Exoplanet Transit Ephemeris Service (http://exoplanetarchive.ipac.caltech.edu/applications/TransitSearch/).  The only midtransit of KOI-351.07 falling within one day of any other transit in the system occurs for KOI-351.06 for Transit 9 (T9), the planet with the shortest period, and thus the one most likely to fall within a day of KOI-351.07 by chance, although its midtransit still falls 17 hours too early.  All times are measured in Barycentric Julian Day (BJD) - 2454833.00 days. We also show the phase of midtransit of KOI-351.07 relative to the period of the other six candidates and demonstrate that this is not a secondary eclipse of another object.  Only the very start of T3's ingress is observed before the end of Q6.  Missing transits T1, T4, and T7, fell into data gaps.  }

\end{deluxetable}
\end{landscape}


\end{document}